\documentclass[english]{article}
\usepackage[T1]{fontenc}
\usepackage[latin9]{inputenc}
\usepackage{geometry}
\usepackage{color}
\usepackage{array}

\makeatletter

\providecommand{\tabularnewline}{\\}

\makeatother

\usepackage{babel}

\begin{document}

\title{Efficient protocols for deterministic secure quantum communication
using GHZ-\emph{like} states }

\maketitle
\begin{center}
Anindita Banerjee$^{1,2}$ and Anirban Pathak$^{1}$
\par\end{center}

\begin{center}
$^{1}$Department of Physics and Materials Science Engineering, 
\par\end{center}

\begin{center}
Jaypee Institute of Information Technology, A-10, Sector-62, Noida,
UP-201307, India. 
\par\end{center}

\begin{center}
$^{2}$ Department of Physics and Center for Astroparticle Physics
and Space Science,
\par\end{center}

\begin{center}
Bose Institute, Block EN, Sector V, Kolkata 700091, India.
\par\end{center}
\begin{abstract}
Two protocols for deterministic secure quantum communication (DSQC)
using GHZ-\emph{like} states have been proposed. It is shown that
one of these protocols can be modified to an equivalent but more efficient
protocol of quantum secure direct communication (QSDC). Security and
efficiency of the proposed protocols are analyzed in detail and are
critically compared with the existing protocols. It is shown that
the proposed protocols are highly efficient. It is also shown that
all the physical systems where dense coding is possible can be used
to design maximally efficient protocol of DSQC and QSDC. Further,
it is shown that dense coding is sufficient but not essential for
DSQC and QSDC protocols of the present kind. We have shown that there
exist a large class of quantum state which can be used to design maximally\textcolor{green}{{}
}efficient DSQC and QSDC protocols of the present kind. It is further,
observed that maximally efficient QSDC protocols are more efficient
than their DSQC counterparts. This additional efficiency arises at
the cost of message transmission rate.
\end{abstract}

\section{Introduction\label{sec:Introduction}}

A protocol for quantum key distribution (QKD) was first introduced
in 1984 by Bennett and Brassard \cite{bb84}. Since then several protocols
for different cryptographic tasks have been proposed. Most of the
initial works \cite{bb84,ekert,b92} on quantum cryptography were
concentrated around QKD. But very soon people realized that quantum-states
can be used for more complex and more specific cryptographic tasks.
For example, In 1999, Hillery \cite{Hillery} had proposed a protocol
for quantum secret sharing (QSS). Almost simultaneously, Shimizu and
Imoto \cite{Imoto} proposed a protocol for direct secure quantum
communication using entangled photon pairs. These protocols had established
that QKD is not essential for secure quantum communication as unconditionally
secure quantum communication is possible without the generation of
keys. Such protocols of direct quantum communications are broadly
divided into two classes: a) Protocols for deterministic secure quantum
communication (DSQC) \cite{dsqc_summation,dsqcqithout-max-entanglement,dsqcwithteleporta,entanglement swapping,Hwang-Hwang-Tsai,reordering1,the:cao and song,the:high-capacity-wstate}
and b) protocols for quantum secure direct communication (QSDC) \cite{ping-pong,vaidman-goldenberg,lm05}.
In DSQC receiver (Bob) can read out the secret message only after
at least one bit of additional classical information for each qubit
is transmitted by the sender (Alice). In contrary, when no such exchange
of classical information is required then the protocols are referred
as QSDC protocols \cite{review}. A conventional QKD protocol generates
the unconditionally secure key by quantum means but then uses classical
cryptographic resources to encode the message. No such classical means
are required in DSQC and QSDC. Further, since all DSQC and QSDC protocols
can be used to generate quantum keys, these protocols of direct communications
are more useful than the traditional QKD protocols. In recent past,
these facts have encouraged several groups to study DSQC and QSDC
protocols in detail {[}\cite{review} and reference there in{]}. 

In the pioneering work of Shimizu and Imoto \cite{Imoto}, they had
cleverly used entangled photon pairs and Bell measurement to achieve
the task of DSQC. In 2002, Beige \emph{et al}. \cite{biege} extended
the idea and proposed another protocol for DSQC using single photon
two-qubit states. But eventually the authors themselves found out
the protocol to be insecure. In the same year, Bostrom and Felbinger
proposed the famous ping-pong protocol of QSDC, which uses EPR states
for communication. This protocol is quasi secure. Such issues with
security of the direct communications are not present in most of the
DSQC and QSDC protocols presented there after. Several unconditionally
secure protocols are presented in last few years. The unconditional
security of those protocols are obtained by using different quantum
resources%
\footnote{In principle the unconditional security arises from quantum non-realism
and conjugate coding.%
}. For example, unconditionally secure protocols are proposed a) with
and without maximally entangled state {[}\cite{dsqcqithout-max-entanglement}
and references there in{]}, b) using teleportation \cite{dsqcwithteleporta}
c) using entanglement swapping \cite{entanglement swapping} d) using
rearrangement of order of particles \cite{reordering1,the:C.-W.-Tsai}
etc. We are specifically interested in the DSQC protocols based on
the rearrangement of orders of particles. Such a protocol was first
proposed by Zhu \emph{et al}. \cite{reordering1} in 2006 but almost
immediately after its publication, it was reported by Li \emph{et
al}. \cite{dsqcqithout-max-entanglement} that the protocol of Zhu
\emph{et al}. is not secure under Trojan horse attack. Li \emph{et
al.} had also proposed a modified version of Zhu \emph{et al}'s protocol.
Thus Li \emph{et al.}'s protocol may be considered as the first unconditionally
secure protocol of DSQC based on rearrangement of order of the particles.
In the last five years, many such protocols of DSQC are proposed.
Very recently, Yuan \emph{et al}. \cite{the:high-capacity-wstate}
and Tsai \emph{et al}. \cite{the:C.-W.-Tsai} have proposed two very
interesting DSQC protocols based on rearrangement of order of the
particles. The Yuan \emph{et al}. protocol uses four-qubit symmetric
$W$ state for communication, while the Tsai \emph{et al.} protocol
utilizes the dense coding of four qubit cluster states. Present work
aims to improve the qubit efficiency of the existing DSQC protocols
and to explore the possibility of designing DSQC and QSDC protocols
using GHZ-\emph{like} states and other quantum states.

GHZ states have been used for quantum information processing since
a long time. Recently, the ideas have been extended to GHZ-\emph{like}
states \cite{Hillery,the:Yang,the:zhang,the:anindita,the:anirban}.
GHZ-\emph{like} states belong to GHZ class and can be generated by
an EPR state, a single qubit state and a controlled-NOT operation.
GHZ-\emph{like} states can be described in general as \begin{equation}
\frac{(|\psi_{i}\rangle\left|0\right\rangle +|\psi_{j}\rangle\left|1\right\rangle )}{\sqrt{2}}\label{eq:a}\end{equation}
where $i,j\epsilon\left\{ 0,1,2,3\right\} $, $i\neq j$ and also
$|\psi_{i}\rangle$ and $|\psi_{j}\rangle$ are Bell states which
are usually denoted as \begin{equation}
\begin{array}{ccc}
|\psi_{0}\rangle & =|\psi_{00}\rangle= & |\psi^{+}\rangle=\frac{\left|00\right\rangle +\left|11\right\rangle }{\sqrt{2}}\\
|\psi_{1}\rangle & =|\psi_{01}\rangle= & |\phi^{+}\rangle=\frac{\left|01\right\rangle +\left|10\right\rangle }{\sqrt{2}}\\
|\psi_{2}\rangle & =|\psi_{10}\rangle= & |\psi^{-}\rangle=\frac{\left|00\right\rangle -\left|11\right\rangle }{\sqrt{2}}\\
|\psi_{3}\rangle & =|\psi_{11}\rangle= & |\phi^{-}\rangle=\frac{\left|01\right\rangle -\left|10\right\rangle }{\sqrt{2}}\end{array}.\label{eq:bell1}\end{equation}

Earlier we have shown that GHZ-\emph{like} states are useful for controlled
quantum teleportation and quantum information splitting \cite{the:anirban}.
Here we have shown that we can form an orthonormal basis set in $2^{3}$
dimensional Hilbert space with 8 GHZ-\emph{like} states, which can
be used for dense coding and DSQC. Thus GHZ-\emph{like} states are
established as a useful resource for quantum information processing.
Remaining part of the paper is organized as follows: In the following
section a protocol for DSQC using GHZ-\emph{like} states without complete
utilization of dense coding is provided. In Section \ref{sec:DSQC_WITH_complete_dc}
we have provided an efficient protocol of DSQC using GHZ-\emph{like}
states with complete utilization of dense coding. In Section \ref{sec:Modification-of-DSQC}
it is shown that the second DSQC protocol (i.e. the one with complete
utilization of dense coding) may be converted to an equivalent QSDC
protocol having better qubit efficiency. In Section \ref{sec:Efficiency-and-security}
we have analyzed the security and efficiency of the proposed DSQC
and QSDC protocols and have shown that the proposed protocols are
unconditionally secure and maximal efficiency can be achieved here.
Finally, we have concluded the work in Section \ref{sec:Conclusions}
and have shown that any set of orthogonal states where dense coding
is possible may be used for DSQC (and QSDC) and consequently for QKD.
We have provided several examples of such quantum states which may
be used for designing of efficient DSQC and QSDC protocols. Further,
it is shown that dense coding is sufficient but not essential for
DSQC and QSDC protocols of the present kind. We have also shown that
there exist a large class of quantum state which can be used to design
maximally\textcolor{green}{{} }efficient DSQC and QSDC protocols of
the present kind.

\section{DSQC using GHZ-\emph{like} states without complete utilization of
dense coding\label{sec:DSQC-without_complete_dc}}

Let us suppose that Alice and Bob are two distant or spatially separated
legitimate/authenticated communicators. Alice wants to transmit a
secret classical message to Bob. The proposed protocol can be implemented
by the following steps:
\begin{enumerate}
\item Without loss of generality we may assume that Alice has prepared $n$
copies of the GHZ-\emph{like} state: \begin{equation}
|\lambda\rangle=\frac{\left|\phi^{+}0\right\rangle +\left|\psi^{+}1\right\rangle }{\sqrt{2}}=\frac{1}{2}\left(|010\rangle+|100\rangle+|001\rangle+|111\rangle\right).\label{eq:lambda}\end{equation}
Now Alice prepares a sequence $P$ of $n$ ordered triplet of entangled
particles as $P=\{p_{1},p_{2}.......,p_{n}\},$ where the subscript
$1,2,...,n$ denotes the order of a particle triplet $p_{i}=\{h_{1},t_{1},t_{2}\},$
which is in the state $|\lambda\rangle$. Symbol $h$ and $t$ are
used to indicate home photon%
\footnote{In the entire manuscript we have used photon and qubit as anonymous
but all the conclusions will remain valid for other form of qubits
too. %
} ($h$) and\textcolor{red}{{} }travel photon ($t$) respectively. 
\item Alice encodes her secret message on sequence $P$ by applying one
of the four two qubit unitary operations $\{U_{00}=X\otimes I,\, U_{01}=I\otimes I,\, U_{10}=I\otimes Z,\, U_{11}=I\otimes iY\}$
on the particles $(h_{1},t_{1})$ of each triplet. The unitary operations
$\{U_{00},\, U_{01},\, U_{10},\, U_{11}\}$ encodes the secret message
$\{00,01,10,11\}$ respectively. Here \begin{equation}
\begin{array}{lcl}
I & = & |0\rangle\langle0|+|1\rangle\langle1|\\
X & = & \sigma_{x}=|0\rangle\langle1|+|1\rangle\langle0|\\
iY & = & i\sigma_{y}=|0\rangle\langle1|-|1\rangle\langle0|\\
Z & = & \sigma_{z}=|0\rangle\langle0|-|1\rangle\langle1|\end{array}.\label{eq:operations1}\end{equation}
These operations $U_{ij}$ $(i,j\epsilon\{0,1\})$ will transform
the GHZ-\emph{like} state $|\lambda\rangle$ into another GHZ-\emph{like}
state $|\lambda_{ij}\rangle,$ where \begin{equation}
\begin{array}{lcl}
|\lambda_{00}\rangle & = & U_{00}|\lambda\rangle=\frac{1}{2}X\otimes I\left(|010\rangle+|100\rangle+|001\rangle+|111\rangle\right)=\frac{1}{2}\left(|110\rangle+|000\rangle+|101\rangle+|011\rangle\right)=\frac{|0\psi^{+}\rangle+|1\phi^{+}\rangle}{\sqrt{2}}\\
|\lambda_{01}\rangle & = & U_{01}|\lambda\rangle=\frac{1}{2}I\otimes I\left(|010\rangle+|100\rangle+|001\rangle+|111\rangle\right)=\frac{1}{2}\left(|010\rangle+|100\rangle+|001\rangle+|111\rangle\right)=\frac{|0\phi^{+}\rangle+|1\psi^{+}\rangle}{\sqrt{2}}\\
|\lambda_{10}\rangle & = & U_{10}|\lambda\rangle=\frac{1}{2}I\otimes Z\left(|010\rangle+|100\rangle+|001\rangle+|111\rangle\right)=\frac{1}{2}\left(-|010\rangle+|100\rangle+|001\rangle-|111\rangle\right)=\frac{|0\phi^{-}\rangle+|1\psi^{-}\rangle}{\sqrt{2}}\\
|\lambda_{11}\rangle & = & U_{11}|\lambda\rangle=\frac{1}{2}I\otimes iY\left(|010\rangle+|100\rangle+|001\rangle+|111\rangle\right)=\frac{1}{2}\left(-|000\rangle+|110\rangle+|011\rangle-|101\rangle\right)=-\frac{\left(|0\psi^{-}\rangle+|1\phi^{-}\rangle\right)}{\sqrt{2}}\end{array}\label{eq:transformed eqns}\end{equation}

\item Alice keeps the home photon $(h_{1})$ of each triplet with her and
prepares an ordered sequence, \\
$P_{A}=\left[p_{1}(h_{1}),p_{2}(h_{1}),...,p_{n}(h_{1})\right]$.
Similarly, she uses all the travel photons to prepare an ordered sequence
$P_{B}=[p_{1}(t_{1},t_{2}),p_{2}(t_{1},t_{2}),...,p_{n}(t_{1},t_{2})]$.
\item Alice disturbs the order of the pair of travel photons%
\footnote{Reordering of sequence of travel photons and reordering of sequence
of unitary operations used to encode the message is equivalent.%
} in $P_{B}$ and create a new sequence\\
 $P_{B}^{\prime}=\left[p_{1}^{\prime}(t_{1},t_{2}),p_{2}^{\prime}(t_{1},t_{2}),...,p_{n}^{\prime}(t_{1},t_{2})\right].$
The actual order is known to Alice only.
\item For preventing the eavesdropping, Alice prepares $m=2n$ decoy photons%
\footnote{In this kind of protocols it is often assumed that number of decoy
photon $m\ll n.$ For example, such an assumption is used in Yuan
\emph{et al}. protocols \cite{the:high-capacity-wstate}. This is
not a correct assumption because if $m\ll n$ then Bob may fail to
detect Eve as there will be a finite possibility that most of the
verification-qubits (decoy states) are not measured by Eve. In contrary,
when $2x$ qubits (a random mix of message qubits and decoy qubits)
travel through a channel accessible to Eve and x of them are tested
for eavesdropping then for any $\delta>0,$ the probability of obtaining
less than $\delta n$ errors on the check qubits (decoy qubits), and
more than $(\delta+\epsilon)n$ errors on the remaining $x$ qubits
is asymptotically less than $\exp[-O(\epsilon^{2}x]$ for large x
\cite{Nielsen chuang-589}. As the unconditional security obtained
in quantum cryptographic protocol relies on the fact that any attempt
of eavesdropping can be identified. Thus to obtain an unconditional
security we always need to check half of the travel qubits for eavesdropping
in other words either half of the entangled states (in our case GHZ-\emph{like}
states) prepared by Alice would be used for checking of eavesdropping
and remaining would be used for encoding or we have to randomly add
decoy qubits whose number would be equal to the total number of travel
qubits that are used for transmission of encoded information. %
}. The decoy photons are randomly prepared in one of the four states
$\{|0\rangle,|1\rangle,|+\rangle,|-\rangle\}$, where $\left|+\right\rangle =\frac{\left|0\right\rangle +\left|1\right\rangle }{\sqrt{2}}$
and $\left|-\right\rangle =\frac{\left|0\right\rangle -\left|1\right\rangle }{\sqrt{2}}$
i.e. decoy photons state is $\otimes_{j=1}^{m}|P_{j}\rangle$, $|P_{j}\rangle\epsilon\{|0\rangle,|1\rangle,|+\rangle,|-\rangle\},$
$(j=1,2,....,m).$ Then Alice randomly inserts these decoy photons
into the sequence $P_{B}^{\prime}$ and creates a new sequence $P_{B+m}^{\prime}$,
which she transmits to Bob. $P_{A}$ remains with Alice%
\footnote{Alternatively Alice may use part of the original sequence for verification
and the remaining part for encoding of information (messaging).%
}.
\item After confirming that Bob has received the entire sequence $P_{B+m}^{\prime}$,
Alice announces the positions of the decoy photons. Bob measures the
corresponding particles in the sequence $P_{B+m}^{\prime}$ by using
$X$ basis or $Z$ basis at random, here $X=\{|+\rangle,|-\rangle\}$
and $Z=\{|0\rangle,|1\rangle\}$. After measurement, Bob publicly
announces the result of his measurement and the basis used for the
measurement. Now Alice has to discard the 50\% cases where, Bob  has
chosen wrong basis. From the remaining outcomes Alice can compute
the error rate and check whether it exceeds the predecided threshold
or not. If it exceeds the threshold, then Alice and Bob abort this
communication and repeat the procedure from the beginning. Otherwise
they go on to the next step.
\item After knowing the position of the decoy photons Bob has already obtained
the sequence $P_{B}^{\prime}$. Now Alice discloses the actual order
of the sequence and Bob uses this information to convert the reordered
sequence $P_{B}^{\prime}$ to the original sequence $P_{B}$. Therefore
Alice needs to exchange $2n$ classical bits. 
\item Alice measures her home photon in computational basis ($Z$ basis)
and announces the result. Bob measures his qubits in Bell basis. From
(\ref{eq:transformed eqns}), it is clear that knowing the results
of measurements of Alice and that of his own measurement, Bob can
easily decode the encoded information. For clarity in Table \ref{table1:dsqc_without}
we have provided a relation between the measurement outcomes and the
secret messages.
\end{enumerate}
\begin{table}
\begin{centering}
\begin{tabular}{|c|c|c|}
\hline 
Alice's measurement result & Bob's measurement result & Decoded secret\tabularnewline
\hline 
0 & $\phi^{+}$ & 01\tabularnewline
\hline 
 & $\phi^{-}$ & 10\tabularnewline
\hline 
 & $\psi^{+}$ & 00\tabularnewline
\hline 
 & $\psi^{-}$ & 11\tabularnewline
\hline 
1 & $\phi^{+}$ & 00\tabularnewline
\hline 
 & $\phi^{-}$ & 11\tabularnewline
\hline 
 & $\psi^{+}$ & 01\tabularnewline
\hline 
 & $\psi^{-}$ & 10\tabularnewline
\hline
\end{tabular}
\par\end{centering}

\caption{Relation between the measurement outcomes and the secret message in
DSQC using GHZ-\emph{like} states without complete utilization of
dense coding.}
\label{table1:dsqc_without}%
\end{table}

An analogous protocol of DSQC based on order rearrangement of particles
have been recently proposed by Yuan \emph{et al.} \cite{the:high-capacity-wstate}.
In their work they have used four qubit symmetric $W$ state for secure
communication of 2 bits of classical information. They have compared
their protocol with an earlier protocol of DSQC introduced by Cao
and Song \cite{the:cao and song}. Cao and Song protocol also uses
4-qubit $W$ state for DSQC but in their protocol each $W$ state
can be used to transmit only one bit of classical information. Keeping
this in mind Yuan \emph{et al.} \cite{the:high-capacity-wstate} claimed
that their protocol has high capacity (compared to Cao and Song protocol)
as each $W$ state can carry two bits of secret information. But it
is straight forward to see that capacity of our protocol is higher
than Yuan \emph{et al.'s} protocol as the state used by us to transmit
two bits of secret message is a 3-qubit GHZ-\emph{like} state. The
advantage is obtained as encoding is done by performing unitary operations
on two qubits (photons) but only one of them is kept as home photon.\textcolor{red}{{}
}The same conclusion would remain valid for GHZ states and any other
tripartite state where dense coding is possible.

Our protocol is considerably more efficient than the protocol of Yuan
\emph{et al.} But the efficiency is not maximal. A maximally efficient
DSQC protocol must be able to transmit $n$ bit secret message using
$n-qubit$ quantum state (or $n$ photons). Such a scheme is recently
introduced by Tsai, Hsieh and Hwang \cite{the:C.-W.-Tsai} by using
4-qubit cluster state. They have used dense coding for the purpose.
In the following section we have shown that it is possible to use
GHZ\emph{-like} state for maximal DSQC.

\section{DSQC using GHZ\emph{-like }states with complete utilization of dense
coding\label{sec:DSQC_WITH_complete_dc}}

In order to establish that a clever use of dense coding may further
increase the efficiency/capacity of the DSQC protocol described in
the last section, first we need to show that dense coding is possible
for GHZ\emph{-like} states. We have established that in the following
subsection and have subsequently described a protocol for DSQC, which
exploit the dense coding.

\subsection{Dense coding using GHZ-\emph{like} states\label{sub:sub_Dense-coding}}

In this sub-section it is shown that GHZ\emph{-like }states can be
used to achieve dense coding. First we need to note that the following
set of 8 GHZ-\emph{like} states are mutually orthogonal: 

\[
\begin{array}{c}
\left\{ \frac{\left|\psi^{+}0\right\rangle +\left|\psi^{-}1\right\rangle }{\sqrt{2}},\,\frac{\left|\psi^{+}0\right\rangle -\left|\psi^{-}1\right\rangle }{\sqrt{2}},\,\frac{\left|\psi^{-}0\right\rangle +\left|\psi^{+}1\right\rangle }{\sqrt{2}},\,\frac{\left|\phi^{+}1\right\rangle -\left|\phi^{-}0\right\rangle }{\sqrt{2}},\right.\\
\left.\frac{\left|\phi^{+}0\right\rangle +\left|\phi^{-}1\right\rangle }{\sqrt{2}},\,\frac{\left|\phi^{+}0\right\rangle -\left|\phi^{-}1\right\rangle }{\sqrt{2}},\,\frac{\left|\phi^{-}0\right\rangle +\left|\phi^{+}1\right\rangle }{\sqrt{2}},\,\frac{\left|\psi^{+}1\right\rangle -\left|\psi^{-}0\right\rangle }{\sqrt{2}}\right\} .\end{array}\]

These states form an orthonormal basis set which may be called GHZ-\emph{like
}basis set%
\footnote{The choice of GHZ-\emph{like} basis set is not unique.%
}. Consequently, these 8 states may be distinguished through measurement
on the GHZ\emph{-like} basis. Assume that initially Alice and Bob
share the following tripartite entangled state of GHZ-\emph{like}
family:

\begin{equation}
{\normalcolor \frac{{\normalcolor {\color{green}{\normalcolor \left|\psi^{+}0\right\rangle +\left|\psi^{-}1\right\rangle }}}}{{\normalcolor {\color{green}{\normalcolor \sqrt{2}}}}}}.\label{eq:initial}\end{equation}
The first two qubits are with Alice and the third one is with Bob.
Now Alice can apply suitable unitary operations (as shown in Table
\ref{table2_densecodingghzlike}) on the first two qubits and transform
the initial state to an orthogonal GHZ-\emph{like} state. Since a
particular GHZ-\emph{like} state can be transformed into all the other
orthogonal GHZ-\emph{like} states. Alice may choose the unitary operations
shown in second column of Table \ref{table2_densecodingghzlike} to
encode $000,001,010,011,100,101,110,111$. After encoding this 3 bit
information, Alice sends her bits to Bob. Now Bob can measure his
qubits in GHZ\emph{-like} basis set and obtain the 3 bit message sent
by Alice. Thus if they share a prior GHZ-\emph{like} entangled state
then Alice can send three bits of classical information by sending
two e-bits. In other words, GHZ\emph{-like} states are useful in dense
coding but they do not provide any advantage over GHZ states. But
our purpose is not to compare the efficiency of dense coding rather
we are interested in maximally efficient DSQC. Let us see how it happens.

\begin{table}[h]
\begin{centering}
\begin{tabular}{|c|c|c|}
\hline 
Message  & Unitary operator  & State\tabularnewline
\hline 
000  & $U_{000}=I\otimes I$  & $\frac{\left|\psi^{+}0\right\rangle +\left|\psi^{-}1\right\rangle }{\sqrt{2}}$\tabularnewline
\hline 
001  & $U_{001}=X\otimes X$  & $\frac{\left|\psi^{+}0\right\rangle -\left|\psi^{-}1\right\rangle }{\sqrt{2}}$\tabularnewline
\hline 
010  & $U_{010}=Z\otimes I$  & $\frac{\left|\psi^{-}0\right\rangle +\left|\psi^{+}1\right\rangle }{\sqrt{2}}$\tabularnewline
\hline 
011  & $U_{011}=iY\otimes I$  & $\frac{\left|\phi^{+}1\right\rangle -\left|\phi^{-}0\right\rangle }{\sqrt{2}}$\tabularnewline
\hline 
100  & $U_{100}=I\otimes X$  & $\frac{\left|\phi^{+}0\right\rangle +\left|\phi^{-}1\right\rangle }{\sqrt{2}},$\tabularnewline
\hline 
101  & $U_{101}=X\otimes I$  & $\frac{\left|\phi^{+}0\right\rangle -\left|\phi^{-}1\right\rangle }{\sqrt{2}}$\tabularnewline
\hline 
110  & $U_{110}=I\otimes iY$  & $\frac{\left|\phi^{-}0\right\rangle +\left|\phi^{+}1\right\rangle }{\sqrt{2}}$\tabularnewline
\hline 
111  & $U_{111}=iY\otimes X$  & $\frac{\left|\psi^{+}1\right\rangle -\left|\psi^{-}0\right\rangle }{\sqrt{2}}$\tabularnewline
\hline
\end{tabular}
\par\end{centering}

\caption{Dense coding using GHZ-\emph{like} states.}
\label{table2_densecodingghzlike}%
\end{table}

\subsubsection{DSQC with the help of dense coding using GHZ-\emph{like} states \label{subsub_:DSQC-with dc}}

The protocol is similar to the protocol introduced in Section 2.1.
with the only difference that Alice does not keep any home photon
and therefore after her encoding she sends the entire state to Bob.
The protocol achieves DSQC in the following steps:
\begin{enumerate}
\item Alice prepares $n$ copies of one of the GHZ-\emph{like} states. Here
we assume that Alice has prepared $n$ copies of the GHZ-\emph{like}
state: \begin{equation}
|\zeta\rangle=\frac{{\normalcolor {\normalcolor {\color{green}{\normalcolor \left|\psi^{+}0\right\rangle +\left|\psi^{-}1\right\rangle }}}}}{{\normalcolor \sqrt{2}}}=\frac{1}{2}\left(|000\rangle+|110\rangle+|001\rangle-|111\rangle\right).\label{eq:lambda-1}\end{equation}
Now Alice prepares a sequence $P$ of $n$ ordered triplet of entangled
particles as $P=\{p_{1},p_{2}.......,p_{n}\},$ where the subscript
$1,2,...,n$ denotes the order of a particle triplet $p_{i}=\{t_{1},t_{2},t_{3}\},$
which is in the state $|\zeta\rangle$. 
\item Alice encodes her secret message on sequence $P$ by applying one
of the eight two qubit unitary operations \[
\begin{array}{c}
\left\{ U_{000}=I\otimes I,\, U_{001}=X\otimes X,\, U_{010}=Z\otimes I,\, U_{011}=iY\otimes I,\right.\\
\left.U_{100}=I\otimes X,\, U_{101}=X\otimes I,\, U_{110}=I\otimes iY,\, U_{111}={\normalcolor {\color{green}{\normalcolor iY\otimes X}}}\right\} \end{array}\]
 on the first two qubits $(t_{1},t_{2})$ of each triplet. The unitary
operations $\{U_{000},\, U_{001},\, U_{010},\, U_{111},U_{100},\, U_{101},\, U_{110},\, U_{111}\}$
encodes $\{000,001,010,011,100,101,110,111\}$ respectively. These
operations $U_{ijk}$ $(i,j,k\epsilon\{0,1\})$ will transform the
state $|\zeta\rangle$ into the state $|\zeta_{ijk}\rangle,$ as shown
in Table \ref{table2_densecodingghzlike}, where \begin{equation}
\begin{array}{lcl}
|\zeta_{000}\rangle & =U_{000}|\zeta\rangle= & \frac{\left|\psi^{+}0\right\rangle +\left|\psi^{-}1\right\rangle }{\sqrt{2}}\\
|\zeta_{001}\rangle & =U_{001}|\zeta\rangle= & \frac{\left|\psi^{+}0\right\rangle -\left|\psi^{-}1\right\rangle }{\sqrt{2}}\\
|\zeta_{010}\rangle & =U_{010}|\zeta\rangle= & \frac{\left|\psi^{-}0\right\rangle +\left|\psi^{+}1\right\rangle }{\sqrt{2}}\\
|\zeta_{011}\rangle & =U_{011}|\zeta\rangle= & \frac{\left|\psi^{-}0\right\rangle -\left|\psi^{+}1\right\rangle }{\sqrt{2}}\\
|\zeta_{100}\rangle & =U_{100}|\zeta\rangle= & \frac{\left|\phi^{+}0\right\rangle +\left|\phi^{-}1\right\rangle }{\sqrt{2}}\\
|\zeta_{101}\rangle & =U_{101}|\zeta\rangle= & \frac{\left|\phi^{+}0\right\rangle -\left|\phi^{-}1\right\rangle }{\sqrt{2}}\\
|\zeta_{110}\rangle & =U_{110}|\zeta\rangle= & \frac{\left|\phi^{-}0\right\rangle +\left|\phi^{+}1\right\rangle }{\sqrt{2}}\\
|\zeta_{111}\rangle & =U_{111}|\zeta\rangle= & \frac{\left|\phi^{-}0\right\rangle -\left|\phi^{+}1\right\rangle }{\sqrt{2}}\end{array}.\label{eq:dsqcby densecoding}\end{equation}

\item Alice disturbs the order of the triplet of travel qubits in $P$ and
create a new sequence \[
P_{B}^{\prime}=\left[p_{1}^{\prime}(t_{1},t_{2},t_{3}),p_{2}^{\prime}(t_{1},t_{2},t_{3}),...,p_{n}^{\prime}(t_{1},t_{2},t_{3})\right].\]
 The actual order is known to Alice only.
\item For preventing from eavesdropping, Alice prepares $m=3n$ decoy photons:
$\otimes_{j=1}^{m}|P_{j}\rangle$, $|P_{j}\rangle\epsilon\{|0\rangle,|1\rangle,|+\rangle,|-\rangle\},$
$(j=1,2,....,m).$ Then Alice randomly inserts these decoy photons
into the sequence $P_{B}^{\prime}$ and creates a new sequence $P_{B+m}^{\prime}$,
which she transmits to Bob. 
\item After confirming that Bob has received the entire sequence $P_{B+m}^{\prime}$,
Alice announces the position of the decoy photons. Bob measures the
corresponding particles in the sequence $P_{B+m}^{\prime}$ by using
$X$ basis or $Z$ basis at random, here $X=\{|+\rangle,|-\rangle\}$
and $Z=\{|0\rangle,|1\rangle\}$%
\footnote{Alternately Alice may announce the position of the decoy states as
well as the basis used to prepare them. In that case Bob would measure
the decoy states using the same basis.%
}. After measurement, Bob publicly announces the result of his measurement
and the basis used for the measurement. Now Alice has to discard the
50\% cases where, Bob has chosen wrong basis. From the remaining outcomes
Alice can compute the error rate and check whether it exceeds the
predecided threshold or not. If it exceeds the threshold then Alice
and Bob abort this communication and repeat the procedure from the
beginning. Otherwise they go on to the next step.
\item After knowing the position of the decoy photons Bob has already obtained
the sequence $P_{B}^{\prime}$. Now Alice discloses the actual order
of the sequence and Bob uses that information to convert the reordered
sequence $P_{B}^{\prime}$ to the original sequence $P$.
\item Now Bob measures his qubits in GHZ\emph{-like} basis (equivalently
he measures any two qubits in Bell basis and the other one in computational
basis). From the third column of Table \ref{table2_densecodingghzlike},
it is clear that from the results of measurements, Bob can easily
decode the encoded information. 
\end{enumerate}
This protocol uses three qubits to communicate three bits of classical
information. This protocol can be extended to any set of states where
all the elements of a basis set are unitarily connected. That means
when you can create the entire basis set by applying unitary operations
on one of the basis vector. This is always true for those states where
dense coding is possible. Consequently our protocol can be extended
to GHZ states and all other states where dense coding is possible.
Some general observations in this regard are noted in Section \ref{sec:Conclusions}.

\section{Modification of the DSQC protocols into the QSDC protocols and their
relevance to QKD\label{sec:Modification-of-DSQC}}

The previous protocol can be easily generalized to a QSDC protocol.
To do so we need to modify Step 3-6 in the above protocol. In the
modified protocol, after Step 2 (i.e. after the encoding is done)
Alice prepares three sequences: $P_{B1}=[p_{1}(t_{1}),p_{2}(t_{1}),....,p_{n}(t_{1})],$
with all the first qubits, $P_{B2}=[p_{1}(t_{2}),p_{2}(t_{2}),....,p_{n}(t_{2})],$
with all the second qubits and $P_{B3}=[p_{1}(t_{3}),p_{2}(t_{3}),....,p_{n}(t_{3})]$
with the remaining qubits. She prepares $3n$ decoy photons as in
Step 4 of the previous protocol and inserts $n$ decoy photons randomly
into each of the three sequences prepared by her. This creates three
extended sequences ($P_{B1+n},\, P_{B2+n},\, P_{B3+n}$) each of which
contain $2n$ qubits. Then she sends the first sequence $P_{B1+n}$
to Bob. After confirming that Bob has received the entire sequence,
she announces the position of decoy photons and checks eavesdropping.
If eavesdropping is found she truncates the protocol otherwise she
sends the second sequence $P_{B2+n}$ to Bob and checks for eavesdropping
and if no eavesdropping is found then she sends the third sequence
and check for eavesdropping. Now Bob can measure the final states
in GHZ-\emph{like} basis and obtain the message sent by Alice. Since
Eve can not obtain more than 1 qubit of a tripartite state (as we
are sending the qubits one by one and checking for eavesdropping after
each step) she has no information about the encoded state and consequently
this direct quantum communication protocol is secure. Thus the rearrangement
of particle order is not required if we do the communication in multiple
steps. Further, since no quantum measurement is done at Alice's end
and rearrangement of particle order is not required, this protocol
does not require any classical communication for the decoding operation.
Thus it is a QSDC protocol. Its efficiency will be naturally higher
than the previous protocol. This is so because here Alice does not
need to disclose the actual sequence and consequently the amount of
classical communication required for decoding of the message is reduced.
But this increase in qubit efficiency is associated with a cost. This
QSDC protocol will be slow as Alice has to communicate in steps and
has to check eavesdropping in the a sequence before she can send the
next sequence. Thus the increase in qubit efficiency will be associated
with a decrease in transmission rate. Following the same path we can
also modify our first protocol and circumvent the use of rearrangement
of particle ordering. That would increase qubit efficiency at the
cost of transmission rate but the protocol would remain a DSQC protocol
as Alice would require to measure the home qubit and communicate the
result (classical communication) to Bob.

Here we would like to note that all DSQC and QSDC protocols can be
reduced to QKD schemes if Alice encodes a random key instead of a
message. Assume that Alice has a random number generator and Alice
encodes the outcome of that in any of the protocols described above.
Following the same protocol Bob will obtain the random key as a message.
Now since both of them have an unconditionally secure quantum key
they may use that for secure communication. Thus all the quantum states
where dense coding is possible may be also used to generate secure
quantum key. From the above discussion, it is straight forward to
note that key generation rate in the above mentioned QSDC protocol
will be less than the corresponding DSQC protocol.

\section{Efficiency and security analysis of the DSQC and QSDC protocols \label{sec:Efficiency-and-security}}

In this section, the security of the proposed DSQC and QSDC protocols
are first analyzed. Then the quantum bit efficiency of the proposed
protocols are analyzed and critically compared with the efficiency
of the existing protocols. It is shown that the proposed protocols
are highly efficient.

\subsection{Security analysis\label{subsub_:Security-analysis}}

The security of the DSQC protocols prescribed above are obtained by
two means. The decoy photon checking technique is used to decide whether
Eve is online or not. Eve can not selectively measure the message
qubits since the decoy photons are randomly inserted in the sequence
$(P_{B+m}^{\prime}$) that is communicated to Bob by Alice. Consequently,
if Eve tries any kind of man-in-the-middle attack he will be disturbing
the decoy qubits and that would lead to the detection of Eve. As the
decoy photons are prepared randomly in $\{|0\rangle,|1\rangle,|+\rangle,|-\rangle\},$
the present security check is of BB-84 type. The same logic ensures
that Eve will be detected in the proposed QSDC protocol too. Here
we need to note an important point, if the number of decoy qubits
are much much less than that of travel qubits used for message encoding
then there is a finite probability that the Eve's man-in-the-middle
attack get undetected. Consequently, Yuan's protocol \cite{the:high-capacity-wstate,Hao},
are insecure and only inclusion of adequate number of decoy photons
can make them secure. Now another important point in this context
is that in addition to have the ability to detect eavesdropping, Alice
and Bob must be able to ensure that the secret message do not leak
to Eve before she is detected. If we only use the decoy qubit technique
then a considerably large amount of information would be leaked to
Eve. For example, in our first protocol of DSQC if Eve attacks all
the transmitting qubits she will obtain 62.5\% of the secret message.
Similarly, she will obtain 100\% of the secret message in our second
protocol of DSQC, and in Yuan's protocol she will obtain 75\% of the
secret message. This can be understood easily. Assume that Eve stores
the sequence $P_{B+m}^{\prime}$ with herself and send a fake sequence
to Bob. Therefore during security check when Alice announces the position
of the decoy qubits, Eve will discard them and measure the rest of
the photon in Bell basis in case of our first protocol and in GHZ-\emph{like}
basis in case of our second protocol. To be precise in case of the
first protocol she may assume that Alice's measurement outcome is
$|1\rangle$ and interpret her Bell measurement results accordingly,
her interpretation will be correct in 50\% cases where Alice's measurement
would actually be $|1\rangle$ and in $\frac{1}{4}th$ of the remaining
cases. Thus 62.5\% of the information would be leaked before detection
of Eve. In case of second protocol Eve will obtain the entire message
and therefore her interpretation will be 100\%. To avoid this potential
attack we need to rearrange the order of the particles. When Alice
disturbs the order of particle sequence then Eve's strategy of intercept-resend
attack can only give her a random sequence of zeroes and ones which
is a meaningless data. In the QSDC protocol proposed above we don't
need to disturb the order of the particles as we check for eavesdropping
after communication of every qubit and Eve can never obtain simultaneous
access to more than one qubit of the entangled state, so she can not
measure the transmitted states in GHZ-\emph{like} basis and consequently
she can not discriminate the encoded states. Further we would like
to note that non-inclusion of the rearrangement of particle ordering
technique makes a few of the existing protocols (e.g.\cite{liu et al})
unsecured under the intercept-resend attack.

\subsection{Efficiency analysis\label{subsub:Efficiency-analysis}}

In the existing literature, two analogous but different parameters
are used for analysis of efficiency of DSQC and QSDC protocols. The
first one is simply defined as \begin{equation}
\eta_{1}=\frac{c}{q}\label{eq:efficency 1}\end{equation}
where $c$ denotes the total number of transmitted classical bits
(message bits) and $q$ denotes the total number of qubits used \cite{Hwang-Hwang-Tsai,the:C.-W.-Tsai}.
This simple measure does not include the classical communication that
is required for decoding of information in a DSQC protocol. Consequently
it is a weak measure. Another measure \cite{dsqcqithout-max-entanglement,liu et al}
that is frequently used and which includes the classical communication
is given as \begin{equation}
\eta_{2}=\frac{c}{q+b}\label{eq:efficiency 2}\end{equation}
where $b$ is the number of classical bits exchanged for decoding
of the message (classical communications used for checking of eavesdropping
is not counted). It is straight forward to visualize that $\eta_{1}=\eta_{2}$
for all QSDC protocols but $\eta_{1}>\eta_{2}$ for all DSQC protocols.
Now while we compare two protocols of DSQC it is important that same
definition of quantum bit efficiency is used. Further, it is important
to note that number of decoy states required to obtain an exponential
security is same as the number of travel qubits (it is neither negligible
as considered in \cite{the:high-capacity-wstate,Hao} nor half of
the total number of qubits) send in message mode (either encoding
is done on these qubits directly or on the qubits entangled to them).
This point is relevant in all such cases where dense coding is not
completely realized and hence all the photons are not send to Bob.
As there is no question of eavesdropping in home photons we don't
need to add decoy photon for them for checking of eavesdropping. This
point is missing in many reported efficiency values. Another important
point is that classical communication required to disclose the actual
order will also contribute to $b.$ This point is not considered in
\cite{the:C.-W.-Tsai} and consequently the efficiency values reported
there are higher than the actual values. Further it is interesting
to note that when we use all the qubits as travel qubits (say when
we use maximal dense coding) then use of decoy photon or part of the
string $P$ for checking of eavesdropping is equivalent but when we
partially use dense coding (or in other words when Alice keeps few
qubits of an $n$-partite entangled state as home qubits and send
the rest as travel qubits) then it is beneficial to use single qubit
decoy photons for eavesdropping checking. This can be understood as
follows: Suppose Alice has $2m$ copies of $2n$-partite entangled
states and she can communicate $x$-bit of classical information using
each $2n$-partite entangled state. Alice use first $n$-particles
of each entangled state as home qubits and sends the remaining $n$-particles
of each sate as travel photon. To check that eavesdropping has not
happened in travel photon either we have to insert $2mn$ decoy photons
or we have to use $m$ copies of entangled states for verification
and remaining for encoding. In the first case we obtain $c=2mx,\, q=2m\times2n+2mn,$
$\eta_{1}=\frac{x}{3n}$ and in the second case $c=mx$ (as only $m$
copies of entangled states are used for encoding), $q=2m\times2n$.
Therefore, $\eta_{1}=\frac{x}{4n}$. This clearly shows that it is
useful to add single photon decoy states in particular cases. We have
already discussed that rearrangement of particle ordering can be avoided
by sending the sequence of qubits in steps and that may convert a
DSQC protocol into a QSDC protocol when all qubits are sent as travel
qubits. In that case efficiency will be $\eta_{1}=\eta_{2}=\frac{1}{2}.$
This imposes an upper-bound on the efficiency of QSDC protocols. This
is so because in the present framework Alice and Bob do not share
any prior entanglement and in such situation one qubit of communication
can not transmit more than one bit of classical information (dense
coding is possible only when prior entanglement exist). Therefore,
$q\geq c$ and $q=c$ iff all the particles of $n-partite$ entangled
states are sent as travel qubits. If these travel qubits are sent
in $n$ steps then we obtain the maximum qubit efficiency $\eta_{2_{max}}=\frac{1}{2}.$
Corresponding rearrangement based single-step DSQC protocols can have
a highest efficiency of $\eta_{2}=\frac{1}{3}$ as we need one bit
of classical communication for each transmitted message qubits (this
is required to disclose the exact sequence). Now the qubit efficiency
of an efficient multi-step DSQC protocol (say instead of using rearrangement
we sent the travel qubits in sequence in our first protocol) will
have a qubit efficiency $\eta_{2}$ greater than their rearrangement
based counter-part but it would be bounded as $\eta_{2}<\frac{1}{2}$.
This can never approach $\eta_{2}=0.5$ as Alice's disclosure of measurement
on home qubits will require some classical communications but it may
reach a value very close to $0.5$ if a $(n-1)$-steps DSQC protocol
is designed with $n-partite$ entangled states having $n\gg1$and
Alice keeps only one qubit of each entangled state as home photon.
In such a situation $\eta_{2}=\frac{n-1}{2(n-1)+1}=\frac{n-1}{2n-1}.$
In Table \ref{table_efficiency} we have provided efficiency of several
existing protocols and have compared them with the proposed protocol.
The comparison clearly establishes that our protocols are highly efficient
and this indicates that GHZ-\emph{like} states can be used as an important
resource for quantum communication. 

\begin{table}
\begin{centering}
\begin{tabular}{|c|>{\centering}p{1in}|>{\centering}p{3cm}|c|}
\hline 
Protocol & Qubit efficiency $\eta_{1}$ in \% & Qubit efficiency~ $\eta_{2}$ in \% & Quantum states used\tabularnewline
\hline 
\cite{Hwang-Hwang-Tsai} & 26.67\% & 22.22\% & Three qubit W state\tabularnewline
\hline 
\cite{the:cao and song} & 16.67\% & 14.29\% & Four qubit W state\tabularnewline
\hline 
 \cite{the:high-capacity-wstate} & 33.33\% & 22.22\% & Four qubit W state\tabularnewline
\hline 
\cite{the:C.-W.-Tsai} & 50\% & 33.33\% & Four qubit cluster state\tabularnewline
\hline 
\cite{liu et al} & 33.33\% & 25\% & Four qubit cluster state\tabularnewline
\hline 
\cite{wang-chinesephys.lett} & 16.67\% & 14.29\% & Four qubit cluster state\tabularnewline
\hline 
 Proposed DSQC without complete & 40\% & 25\% & Three qubit GHZ-\emph{like} state\tabularnewline
utilization of densecoding &  &  & \tabularnewline
\hline 
 Proposed DSQC with complete & 50\% & 33.33\% & Three qubit GHZ-\emph{like }state\tabularnewline
utilization of densecoding &  &  & \tabularnewline
\hline 
Proposed QSDC protocol & 50\% & 50\% & Three qubit GHZ-\emph{like} state\tabularnewline
\hline
\end{tabular}
\par\end{centering}

\caption{Comparison of quantum bit efficiency of different protocols.}
\label{table_efficiency}%
\end{table}

\section{Conclusions\label{sec:Conclusions}}

Two protocols of DSQC using tri-partite GHZ-\emph{like} states have
been proposed. One of them is modified to construct a 3-step QSDC
protocol. It is shown that these protocols are efficient and secure.
It is also shown that some of the existing protocols are insecure
and qubit efficiency claimed in some other cases are more than their
actual efficiency. For example, Yuan et al \cite{the:high-capacity-wstate,Hao}
protocol is insecure under intercept resend attack as the number of
decoy qubits are negligible compare to the total number of travel
qubits used for encoding of the secret message. On the other hand,
the qubit efficiency computed in \cite{the:C.-W.-Tsai} is an over
estimate as they have not considered the classical communication required
for disclosure of actual order of the particles. We have reported
corrected efficiency of all such protocols and have established that
the GHZ-\emph{like} states can achieve maximum qubit efficiency. It
established GHZ-\emph{like} states as an important resource for quantum
communication but it does not indicate any advantage of these states
over the other states. To be precise, a large number of protocols
of DSQC \cite{the:C.-W.-Tsai,the:high-capacity-wstate,review} have
been reported in recent past. All these protocols, which use reordering
of the sequence of particles, decoy photon and entangled channel,
are essentially same. The schemes which do not use dense coding directly,
for example, the protocol described in \cite{the:high-capacity-wstate}
and the analogous scheme described here in Section \ref{sec:DSQC-without_complete_dc}
do not claim to use dense coding in true sense but what it does is
simply partial utilization of dense coding. From these facts one may
conclude that dense coding is essential for efficient DSQC protocols.
Here we will show that dense coding is sufficient for designing of
a maximally efficient DSQC protocol in an analogy to our second protocol
and can be modified to a multi-step QSDC protocol by the same manner
as we have done here but dense coding is neither essential for the
efficient DSQC nor for the QSDC protocols. 

Let us assume that we have a set $Q=\{Q_{0},Q_{1},......,Q_{2^{n}-1}\}$
of $n-$partite orthonormal vectors which spans the $2^{n}$ dimensional
Hilbert space. Further we assume that these state vectors can be unitarily
transformed to each other. In other words there exist a set of unitary
operations $U=\{U_{0}^{i},U_{1}^{i},....U_{2^{n}-1}^{i}:U_{j}^{i}Q_{i}=Q_{j}\}$
such that the unitary operations can transform a particular element
$Q_{i}$ of set $Q$ into all the other elements of set $Q$. Now
suppose Alice prepares multiple copies of the state vector $Q_{i}$.
She can encode $n$-bit message by using an encoding scheme in which
$\{U_{0}^{i},U_{1}^{i},\cdots,U_{2^{n}-1}^{i}\}$ are used to encode
$\{0_{1}0_{2}\cdots0_{n},\,0_{1}0_{2}\cdots1_{n},\cdots,1_{1}1_{2}\cdots1_{n}\}$
respectively. If these encoded messages are sent to Bob then Bob can
unambiguously decode the message since the states received by him
are mutually orthogonal. Security can be achieved by insertion of
decoy qubits and either by rearrangement of the particle order or
by sending the encoded steps in multiple states. This is sufficient
for construction of an efficient DSQC or QSDC scheme and it does not
require dense coding. 

Dense coding is a special case of the above idea. To be precise, dense
coding is possible if and only if $U_{j}^{i}$ are $m$ qubit operators,
where $m<n$. In a good dense coding protocol the operators are chosen
in such a way that $m=\frac{n}{2}$ for even $n$ and $m=\frac{n}{2}+1$
for odd $n$. Now for dense coding, Alice and Bob shares a quantum
channel $Q_{i}$ in such a way that $m$ qubits of $Q_{i}$ are with
Alice and the remaining $n-m$ qubits are with Bob, who knows the
initial state $Q_{i}$ prepared by Alice. Alice may encode a $n$-bit
message by operating any of $2^{n}$ unitary $m$-qubit operators
available with her (say she applies $U_{J}^{i}$) and send her photons
to Bob. Now since the orthonormal states are distinguishable, Bob
can measure his qubits (in Q basis) and find the state $Q_{j}$. Since
he already knows that the initial state was $Q_{i}$ , now he knows
that Alice has send him a bit string indexed as $j$. Thus $n$-bit
information are send with $m<n$ qubits. This is the essence of dense
coding. This does not involve any security measures. If we just need
to modify it for DSQC then we assume that Alice and Bob do not share
any entanglement. Alice prepares $n$ copies of $Q_{i}$, prepares
a sequence, apply unitary operations as per the secret message she
wants to send, changes its order, inserts decoy photons and sends
the entire sequence to Bob. Now first Bob confirms that he has received
a sequence of appropriate length, then Alice announces the positions
of the decoy photons, Bob measures them and announces the result and
the basis used to measure them. Alice uses that result to compute
the error rate to detect Eve. If the  error rate is below the threshold,
Alice discloses the sequence, Bob reorders his sequence and measures
it in $Q$ basis to know the secret send by Alice. In this entire
protocol no specific channel is used. This is true in general and
consequently any existing dense coding protocol can be used for DSQC.
Further, Bob will be able to distinguish among $2^{n}$ orthogonal
states iff he has access to all the $n$ qubits and consequently Alice
has to send the entire sequence to Bob. In all such cases quantum
bit efficiency $\eta_{2}$ will be maximum. No quantum channel $(Q_{i})$
has any advantage over the others ($Q_{j\neq i}$). Since dense coding
has been already reported in several systems, the present discussion
implies that the maximally efficient DSQC (and their maximally efficient
multi-step QSDC counterparts) can be achieved with large number of
alternatives. For example. dense coding is recently reported in different
classes of genuine quadripartite entangled states \cite{Pati}, such
as $\left|\Omega\right\rangle =\frac{1}{\sqrt{2}}\left(\left|0\right\rangle \left|\varphi^{+}\right\rangle \left|0\right\rangle +\left|1\right\rangle \left|\varphi^{-}\right\rangle \left|1\right\rangle \right)$,~$\left|Q_{4}\right\rangle =\frac{1}{2}\left(\left|0000\right\rangle +\left|0101\right\rangle +\left|1000\right\rangle +\left|1110\right\rangle \right)$,
$\left|Q_{5}\right\rangle =\frac{1}{2}\left(\left|0000\right\rangle +\left|1011\right\rangle +\left|1101\right\rangle +\left|1110\right\rangle \right)$.
All these states can now be used for efficient DSQC. 

Thus dense coding is sufficient to construct efficient DSQC protocols
and several alternate quantum channel exist. But as dense coding is
not essential for DSQC, there exist a large number of quantum states
which does not show maximal dense coding (or which does not show dense
coding at all) but which may be used for designing of efficient protocol
of DSQC and QSDC. To be precise, if there exist a set of orthonormal
states for which $m=n$, then dense coding would not be possible but
DSQC would be possible. Examples of such a system are tripartite and
quadripartite $W$ states. In brief, we have numerous alternative
quantum states which may be used for DSQC and at the end of the day
it appears to be matter of convenience of the experimentalist. Whichever,
set of orthogonal states $Q$ and unitary operations $U$ will be
easier to generate experimentally will be more useful for the realization
of DSQC. This is so because quantum bit efficiency $\eta_{2}$ is
same for all these systems. Further, we would like to note that since
the efficiency would remain same, it does not really make sense to
investigate the possibility of DSQC using newer systems involving
more complex quantum states (say 6 qubit $W$ state or Brown state).
Here we have exploited the intrinsic symmetry of the existing protocols
and have used simple minded logic to reach at strong conjecture that
no new and complex system will provide better quantum bit efficiency
as long as present class of protocols are concerned.

Further, we wish to add that Werner \cite{Werner} had shown that
there exists a one to one correspondence between dense coding and
teleportation when these schemes are assumed to be tight. Here we
have shown that there exists a one to one mapping between dense coding,
DSQC and QSDC. Consequently all the quantum states used for perfect
teleportation can be used for DSQC, QSDC and QKD. These simple but
interesting observations open up the window for experimentalist to
use different experimentally realizable quantum states (as per their
convenience to create it experimentally and to maintain it) for the
purpose of QKD, QSDC and DSQC.

~

\textbf{Acknowledgment: }AP thanks Institute of Mathematical Sciences,
Chennai, India (IMSc) for the warm hospitality and facilities provided
during his visit. Part of this work is done in IMSc. He also thanks
Department of Science and Technology (DST), India for support provided
through the DST project No. SR/S2/LOP/2010.


\begin{thebibliography}{29}
\bibitem{bb84}C. H. Bennett and G. Brassard, in Proceedings of the
IEEE International Conference on Computers, Systems, and Signal Processing,
Bangalore, p. 175 (1984).

\bibitem{ekert}A.K. Ekert, Phys. Rev. Lett. \textbf{67,} (1991) 661.

\bibitem{b92}C.H. Bennett, Phys. Rev. Lett. \textbf{68,} (1992) 3121.

\bibitem{Hillery}M. Hillery, V. Buzek and A. Bertaiume, Phys. Rev.
A \textbf{59,} (1999) 1829.

\bibitem{Imoto}K. Shimizu and N. Imoto, Phys. Rev. A \textbf{60,}
(1999), 157.

\bibitem{dsqc_summation}J. Liu \emph{et al.}, Chin. Phys. Lett. \textbf{23,}
(2006) 2652.

\bibitem{dsqcqithout-max-entanglement}X. H. Li \emph{et al}., J.
Korean Phys. Soc. \textbf{49,} (2006) 1354. 

\bibitem{dsqcwithteleporta}F. L. Yan and X. Zhang, Euro. Phys. J.
B\textbf{ 41,} (2004) 75.

\bibitem{entanglement swapping}Z. X. Man, Z. J. Zhang and Y. Li,
Chin. Phys. Lett. \textbf{22,} (2005) 18.

\bibitem{Hwang-Hwang-Tsai}T. Hwang, C. C. Hwang and C W Tsai, Euro.
Phys. J. D \textbf{61,} (2011) 785.

\bibitem{reordering1}A. D. Zhu, Y. Xia, Q. B. Fan, and S. Zhang,
Phys. Rev. A \textbf{73,} (2006) 022338.

\bibitem{the:cao and song}H. J. Cao and H.S. Song, Chin. Phys. Lett.
\textbf{23,} (2006) 290.

\bibitem{the:high-capacity-wstate}H. Yuan \emph{et al.}, Int. J.
Theo. Phys. \textbf{50}, (2011) 2403. 

\bibitem{ping-pong}K. Bostrom and T. Felbinger, Phys. Rev. Lett.
\textbf{89}, 187902 (2002).

\bibitem{vaidman-goldenberg}L. Goldenberg and L. Vaidman, Phys. Rev.
Lett. \textbf{75,} (1995) 1239.

\bibitem{lm05}M. Lucamarini and S. Mancini, Phys. Rev. Lett. \textbf{94},
140501 (2005).

\bibitem{review}G. Long, \emph{et al}., Front. Phys. China, \textbf{2}
(2007) 251.

\bibitem{biege}A. Beige, B. G. Englert, C.Kurtsiefer and H.Weinfurter,
Acta Phys. Pol. A \textbf{101,} (2002) 357. 

\bibitem{the:C.-W.-Tsai}C.W. Tsai, C.R. Hsieh, and T. Hwang, Eur.
Phys. J. D \textbf{61,} (2011) 779.

\bibitem{the:Yang}Yang \emph{et al.}, Int. J. Theo. Phys.\textbf{
48,} (2009) 516.

\bibitem{the:zhang}Q.Y. Zhang \emph{et al.}, Int. J. Theo. Phys.
\textbf{48,} (2009) 3331.

\bibitem{the:anindita}A. Banerjee, K. Patel and A. Pathak, Int. J.
Theo. Phys. \textbf{50,} (2011) 507. 

\bibitem{the:anirban}A. Pathak and A. Banerjee, Int. J. Quant. Info.
\textbf{9} (2011) 389.

\bibitem{Nielsen chuang-589}M. A. Nielsen and I. L. Chuang, Quantum
Computation and Quantum Information, Cambridge University Press, New
Delhi (2008) 589.

\bibitem{Hao}YUAN Hao \emph{et al.}, Commun.Theor. Phys.\textbf{
55,} (2011) 984.

\bibitem{liu et al}W. J. Liu \emph{et al}. Chin.Phys. B \textbf{18,}
(2009) 4105\textcolor{green}{.}

\bibitem{wang-chinesephys.lett}G. Y. Wang, X. M. Fang and X. H. Tan,
Chin. Phys. Lett. \textbf{23,} (2006) 2658. 

\bibitem{Pati}B. Pradhan, P. Agrawal and A. K. Pati, quant-ph/0705.1917.v1
(2007).

\bibitem{Werner}R. F. Werner, J. Phys. A: Math. Gen. \textbf{34,}
(2001) 7081. 
\end{thebibliography}
\end{document}